\documentclass[12pt,preprintnumbers]{revtex4}
\usepackage{epsfig}

\def\eq#1{(\ref{#1})}
\def\fig#1{{Fig.~\ref{#1}}}
\def\re#1{{Ref.~\cite{#1}}}
\def\order#1{\mathcal{O}{(#1)}}
\newcommand{\beq}{\begin{equation}}
\newcommand{\eeq}{\end{equation}}
\newcommand{\beqar}[1]{\begin{eqnarray}\label{#1}}
\newcommand{\eeqar}{\end{eqnarray}}
\newcommand{\as}{\alpha_s}

\newcommand{\lag}{\mathcal{L}}

\newcommand{\vac}{|\epsilon_\mathrm{v}|}

\def\npb#1#2#3{    {\it Nucl. Phys. }{\bf B#1} (#2) #3}

\def\plb#1#2#3{    {\it Phys. Lett. }{\bf B#1} (#2) #3}

\def\sjnp#1#2#3{   {\it Sov. J. Nucl. Phys. }{\bf #1} (#2) #3}

\def\jetpl#1#2#3{  {\it JETP Lett. }{\bf #1} (#2) #3}

\begin{document}

\preprint{BNL-NT-04/9}
\preprint{TAUP-2766-04}

\title {QCD in curved space-time: a conformal bag model} 


\author{Dmitri Kharzeev}
\affiliation{ Physics Department, Brookhaven National Laboratory,\\
Upton, NY 11973-5000, USA}

\author{Eugene Levin}
\affiliation{HEP Department, School of Physics and Astronomy\\
Tel-Aviv University, Ramat Aviv, 69978, Israel} 

\author{Kirill Tuchin}
\affiliation{Physics Department, Brookhaven National Laboratory,\\
Upton,\mbox{} NY 11973-5000, USA}

\date{ \today}

\begin{abstract}

We construct an effective low energy Lagrangian 
of gluodynamics which 
(i) satisfies all constraints imposed by the Renormalization Group; (ii) is scale and 
conformally invariant in the limit of vanishing vacuum energy density;
(iii) matches onto the perturbative theory at short distances. 
This effective theory has a dual description as   
classical gluodynamics on a curved 
conformal background. Color fields are dynamically confined, and 
the strong coupling freezes at distances larger than 
the glueball size. We also make specific predictions (in particular, on the $N_c$ dependence 
of glueball properties) which can be tested in lattice 
simulations of gluodynamics.  

\end{abstract}

\maketitle


\section{Introduction}

\subsection{Instability of the perturbative vacuum}

It has been known for a long time that the perturbative QCD vacuum is
not the true vacuum of the theory. 
One way to see this is to examine the 
derivation of the asymptotic freedom \cite{GW} in the effective potential 
method.  
The real part of the one-loop potential of gluodynamics for a constant 
chromomagnetic field $H$ reads 
\cite{masa} 
\beq\label{oneloop}
\mathrm{Re}\, V_\mathrm{pert}(H)\,=\, \frac{1}{2}\, H^2\,+\,(g\, H)^2\,
\frac{b}{32\,\pi^2}\,\left(\ln\frac{g\,H}{\mu^2}\,-\,\frac{1}{2}
\right),
\eeq   
where $\mu$ is the renormalization scale, and $b=11\, N_c/3$.
 This potential has a minimum at $H=H_0$:
(\ref{oneloop})
\beq\label{H0} 
g\,H_0\,=\,\mu^2\,e^{-\frac{16\,\pi^2}{b \,g^2(\mu)}}.
\eeq
which is natural to identify with the perturbative vacuum. 

However it was soon realized that this perturbative vacuum is unstable. 
It is instructive to trace the origin of this instability in the effective potential method, which was pointed out in Ref.\cite{no}.  
Consider the Landau levels
of a particle of spin $s$ and four-momentum $p$ 
in a constant chromomagnetic field $H \hat z$ \cite{tsai}:
\beq\label{lan}
p^\mu\,p_\mu = 2\,g\,H\,(n\,+\,1/2)-2\, s_z\, g\, H,
\eeq
where $s_z$ is a projection of the spin on the direction of the
chromomagnetic field. The effective potential \eq{oneloop} can be
calculated as \cite{no}
\beq\label{summ}
V_\mathrm{pert}(H)\,=\,\frac{g\,H}{4\,\pi^2}\,
\int\,dp_z\,\sum_{n=0}^\infty\,\sum_{s_z=\pm 1}\,
\sqrt{2\,g\,H\,(n+1/2-s_z)+p_z^2}.
\eeq
Its real part yields \eq{oneloop}, while the imaginary part can be
calculated as 
\beq\label{imag}
\mathrm{Im}\,V_\mathrm{pert}(H)\,=\,\frac{g\,H}{4\,\pi^2}\,
\int_{-\,g\,H}^{+\,g\,H}dp_z\,\sqrt{p_z^2-g\,H-i\,0}\,=\,
-\,\frac{g^2\,H^2}{8\,\pi}.
\eeq
Therefore, the perturbative vacuum corresponding to the minimum of the
perturbative potential \eq{H0} is unstable. The instability is caused by
the mode $n=0$ and $s_z=1$ (spin direction parallel to the field);
note that $n=0$ corresponds to Landau level of the largest radius $\sim
1/\sqrt{g H}$, i.e. to the infrared region of the theory.  This means that
perturbative QCD is ill-defined at large distances \cite{Gribov}, and we 
may have to
describe the theory in terms of other variables.

\subsection{QCD in a cavity}

The breakdown of the perturbative approach (at least, at the one--loop
level) has to happen at some critical value of the chromomagnetic field
$H_c > H_0 = \Lambda_{QCD}^2/g$. This means that weaker color fields
cannot penetrate the physical vacuum, and the necessary condition for the
applicability of the perturbative approach is that the energy density of
the color field is sufficiently high:
\beq 
\epsilon_H = \frac{H^2}{8 \pi} > \frac{ \Lambda^4_{QCD}}{32 \pi^2 \alpha_s}
\label{penet}
\eeq

The condition (\ref{penet}) means that the color fields can be properly
defined only at distances smaller than $R_\mathrm{conf} \sim
\Lambda^{-1}_{QCD}$. For perturbative theory to make sense, it has
therefore be constrained within a  cavity of radius $R_\mathrm{conf}$, 
with appropriate boundary conditions. A
possible realization of this idea is the MIT bag model \cite{MITbag} 
where the colored fields are required to vanish at the surface of a 
sphere.
 
 It is well known that the presence of 
 boundary conditions leads to the emergence of Casimir vacuum energy 
 $\epsilon_{C} \sim R^{-4}_\mathrm{conf} \sim \Lambda^{4}_{QCD}$; in MIT 
bag
model, it is represented by a ``bag constant".
 It is also known that a theory in flat space-time
 in the presence of non-trivial boundary conditions can often be
conveniently described as a theory in a curved background
 \cite{BD}.
 In this paper we will argue that such a description is possible for
gluodynamics. We develop an effective theory
 which has the following \emph{dual descriptions}: (i) classical  
gluodynamics in a curved conformal space-time background and (ii)
 gluodynamics in flat space-time coupled to scalar glueballs (which in
this case play the role of dilatons saturating the
 correlation functions of the trace of the energy-momentum tensor).
 The representation of the effective theory in flat space-time appears
quite similar to the non-topological soliton model
 of Friedberg and Lee \cite{FL} which describes quarks coupled to a scalar
self-interacting field, and more generally to the approach outlined in Ref. \cite{LW}; we will return
 to the discussion of this topic later.

 It is clear that such an approach should have its limitations. Consider,
for example, the dependence on the number of colors
 $N_c$: the energy density of the gluon field $\epsilon_H \sim (N_c^2 -1)
$, whereas the Casimir vacuum energy
 $\epsilon_{C} \sim \Lambda^{4}_{QCD} \sim N_c^0$.  One therefore can
expect that the effect of the boundary will
 diminish at large $N_c$, and so the approach may not have a smooth $N_c
\to \infty$ limit.
 
\subsection{Renormalization group and low energy theorems}

The basic property of the perturbative effective potential
\eq{oneloop} is its invariance under the Renormalization Group (RG)
transformations. We would like to preserve this fundamental property
at all distances \cite{Pagels,ms,schechter}. For this purpose, we need to encode 
the properties of RG in a set of low energy theorems for the correlation functions 
of the trace of the energy momentum tensor.

Let us sketch the derivation of these theorems, as they represent  the guiding principle for 
the construction of our effective theory. 
Consider an expectation
value of an operator $O$ of canonical dimension $d$; it can be written
down as 
\beq\label{sred}
\langle O\rangle\,\sim\, \left[ M_0\,e^{-\,\frac{8\,\pi}
{b\, g^2(\mu)}}\right]^d.
\eeq
On the other hand, the dependence of the QCD Lagrangian on the
coupling is 
\beq\label{class}
\lag_\mathrm{QCD}\,=\,(-1/4g^2)\tilde F^{a\mu\nu}\tilde F^a_{\mu\nu},
\eeq
where $\tilde F = g F$ is the rescaled gluon field. Following
\re{svz,svz1} we can write down the expectation value of the operator
$O$ in the form of the functional integral and  differentiate with
respect  to $  (-1/4g^2(\mu))$ to get
\beq\label{prt}
i\,\int\,dx\,\langle T\{\,O(x)\,,\, \tilde F^2(0)\,\}\rangle\,=\,
-\frac{d}{d(-1/4g^2)}\,\langle O\rangle . 
\eeq
Combining \eq{sred} and \eq{prt} we obtain the relation
\cite{svz,svz1}
\beq\label{LET0}
i\,\lim_{q\rightarrow 0}\,\int dx\,e^{i\,q\,x}\,\langle0|T\{\,O(x)\,,\,
\frac{\beta(\as)}{4\,\as}\, F^2(0)\,
\}|0\rangle_\mathrm{connected}=
\langle O\rangle\, (-4)\,.
\eeq
This expression can be easily iterated by consequent differentiation 
like in \eq{prt} to obtain a set of relations
between Green's functions involving an arbitrary number of operators $F^2$. 
We can rewrite those relations using expression for the scale anomaly
in QCD in terms 
of the trace of the energy momentum tensor $\theta_\mu^\mu$ ($d=4$)
\beq\label{anom}
\theta_\mu^\mu\,=\,\frac{\beta(g)}{2\,g}\,
F^a_{\mu\nu}\, F^{a\mu\nu}.
\eeq
Substituting also $\theta_\mu^\mu$ for $O$ we obtain 
the following set of low energy theorems
for different Green's functions involving operator
$\theta_{\mu}^{\mu}(x)$:
\beq\label{LET}
i^n
\int dx_1\ldots dx_n\,\langle0|T\{\theta_{\mu_1}^{\mu_1}(x_1),
\ldots,\theta_{\mu_n}^{\mu_n}(x_n),\theta_{\mu}^{\mu}(0)
\}|0\rangle_\mathrm{connected}\,=\,
\langle\theta_{\mu}^{\mu}(x)\rangle\, (-4)^n.
\eeq
Eqs.~\eq{LET0},\eq{LET} show that although
the scale symmetry of the classical Yang-Mills \eq{class} has
been broken down by quantum fluctuations \cite{sa}, there still remains a
symmetry imposed by the invariance of the observables 
under the Renormalization Group. 
In the next section we are going to
construct an effective Lagrangian which saturates  \eq{LET} at long
distances and matches onto the one-loop perturbative effective
Lagrangian at short distances.
  
\section{Effective Lagrangian}

We start with the derivation of the effective Lagrangian using the
mathematical trick suggested in \cite{ms}. Consider the Yang-Mills
theory on a curved conformally flat background in $d$ dimensions. 
The background is given by the metric
\beq\label{metric}
\mathrm{g}_{\mu\nu}(x)\,=\,e^{h(x)}\, \delta_{\mu\nu},
\eeq
and the action by
\beq\label{action}
S\,=\,-\frac{1}{4\, g^2}\,\int d^dx\,\sqrt{- \mathrm{g}} \,
\mathrm{g}^{\mu\nu}\, \mathrm{g}^{\lambda\sigma}\, \tilde F^a_{\mu\lambda}\,
\tilde F^a_{\nu \sigma}.
\eeq
where $\mathrm{g}=\det \mathrm{g}_{\mu\nu}$. Recall that the classical 
Yang-Mills Lagrangian in flat spacetime is scale and conformally
invariant only in four dimensions. On the contrary, it can be proven 
\cite{ms} that the theory on the curved background given by \eq{metric},\eq{action} is scale
and conformally invariant in any number of dimensions $d$ -- 
 this means that regularization does not bring into the theory
\eq{action} any dimensionful parameters. Upon regularization the action
\eq{action} acquires an additional term in $d=4$:
\beq\label{addterm}
\Delta S\,=\,-\frac{1}{4\, g^2}\, \int d^4x\, e^{2h}\,
\left[-\frac{b\,g^2}{32\,\pi^2}(\tilde F^a_{\mu\nu})^2\right].
\eeq  
The effective one-loop action of Yang-Mills field in the external constant  
conformally flat gravitational field is given by the sum of
\eq{action} and \eq{addterm}; it is obviously scale and conformally invariant.
The term \eq{addterm} corresponds to the anomalous 
second term in the right hand side of \eq{oneloop}. Therefore, the 
scale anomaly of QCD manifests itself in the theory defined by  \eq{metric} and \eq{action}
through a term containing the axillary scalar field $h$ \cite{ms}, without any dimensionful parameters. 
In a dual, and more conventional,  flat space-time description the scale anomaly exhibits itself in the phenomenon of dimensional 
transmutation, which brings in a dimensionful parameter explicitly. 

The kinetic part for the field $h(x)$ can be obtained in a manifestly
scale and conformally invariant way using the Einstein-Hilbert
Lagrangian for the one-loop effective Yang-Mills field
\beq\label{act}
S\,=\,\int d^4x\,\sqrt{-\mathrm{ g}}\left( \frac{1}{8\,\pi\, G}\, R
 \,-\frac{1}{4\, g^2} \,
\mathrm{g}^{\mu\nu}\, \mathrm{g}^{\lambda\sigma}\, \tilde F^a_{\mu\lambda}\,
\tilde F^a_{\nu \sigma}\,-\, 
e^{2h}\,\theta_\mu^\mu
\right),
\eeq
where $R$ is the Ricci scalar and $G$ is some dimensionful constant;  
we substituted \eq{anom} into
the square brackets of \eq{addterm}. We can now  use a
well-known expression for the Riemann tensor $R_{\mu\nu}$ \cite{ll}  to write
down the dynamical terms for the field $h(x)$ which obey the scale and
conformal symmetry. Using \eq{metric} we get  
\beq\label{riemann}
R\,\sqrt{-\mathrm{g}}\,\equiv\,R_\mu^\mu\,\sqrt{-\mathrm{g}}
\,=\,e^h\,\frac{3}{2}\,(\partial_\mu h)^2.
\eeq 
Note that by writing \eq{riemann} we explicitly neglected terms of
higher order in derivatives and constrained ourselves to the
Einstein's gravity.  This correspond to an expansion in powers of a slowly
varying background field.

The vacuum expectation value of the energy-momentum tensor reads
\beq\label{treb}
\langle\theta_\mu^\mu\rangle\,=\,-\,4\,\vac.
\eeq 
The perturbative contribution to \eq{treb} is given by
\eq{anom}. Since the perturbative vacuum \eq{H0} is not stable, 
it is natural to assume that the dominant contribution to the energy density of the physical
vacuum comes from non-perturbative modes. It is therefore 
convenient to separate the perturbative
contribution to the $\theta_\mu^\mu$ in the following way: 
\beq\label{split}
\theta_\mu^\mu\,=\, \theta_\mu^\mu(\mathrm{pert.})\,-\,4\,\vac;
\eeq
(we will argue below (see \eq{min}) that the physical vacuum is indeed
independent of the value of the external chromomagnetic field.)
Combining \eq{act},\eq{riemann} and \eq{split} we arrive at the
expression for the effective
one-loop action in the conformally flat gravitational field 
\beq\label{act5}
S\,=\,\int d^4x \left(
\frac{4\,\vac}{m^2}\,e^h\,\left(\partial_\mu\,h\right)^2\,-\,
\frac{1}{4}\,(F_{\mu\nu}^a)^2\,+\,\vac\, e^{2\,h}\,-\,
\frac{1}{4}\,e^{2\,h}\,\left[-\frac{b\,g^2}{32\,\pi^2}\,
  (F_{\mu\nu}^a)^2\right]\right),
\eeq
where the new dimensionful constant $m$ was introduced instead of $G$
\cite{klt}.

At this point it is important to note that one can easily read the running
coupling constant off \eq{action} and \eq{addterm}. It can be seen that
$-e^{2h}$ plays a role of the familiar perturbative logarithm
$2\ln(q^2/\mu^2)$. Hence our effective theory is applicable when
$q^2<\mu^2$. In the infrared region the perturbative expressions break
down. However it is possible to remove the explicit dependence on  
 the strong coupling from the effective
action by performing the following redefinition in \eq{act5}
\beq\label{transf}
h\,\rightarrow\, h\,-\,2\,\ln\varsigma\,,\quad
\vac\,\rightarrow\, \varsigma^4\,\vac\, ,\quad
m^2\,\rightarrow\,\varsigma^2\, m^2\,,
\eeq
where $\varsigma^4=g^2\,b/32\,\pi^2$. Eq.~\eq{transf} is just a change  
of mass unit.
  
Finally, we have to perform Legendre transformation of the action
\eq{act5} to get the minimum of the effective potential at the minimum
of the field $\chi$ which is canonically conjugated to the field
$h$ \cite{ms}. The result reads \cite{klt}
\beq\label{LAGR}
\lag \,=\, \frac{\vac}{m^2}\,\frac{1}{2}\,e^{\chi/2}\, (\partial_{\mu}\chi)^2\,
+\,\vac\, e^\chi\,(1-\chi)\, -\,\frac{1}{4}\,(F^a_{\mu\nu})^2\,
+\,e^{\chi}\,(1-\chi)\,\frac{1}{4}\,
\,(F^a_{\mu\nu})^2,
\eeq
where we included the factor $\sqrt{-\mathrm{g}}$ in the definition of
$\lag$. The Lagrangian \eq{LAGR} defines our effective theory. At large distances \eq{LAGR} reduces to the low energy effective
Lagrangian of Refs.~\cite{schechter,ms}.
Indeed, the Yang-Mills action 
is scale invariant in the external field $h$, which 
implies that $F_{\mu\nu}^a\sim q^2\rightarrow 0$ at long distances. 
We will investigate the region of applicability of the effective
Lagrangian \eq{LAGR} in the following sections.

The mathematical trick of putting the theory in curved space-time background 
which we used in derivation of \eq{LAGR} 
gives a simple way to keep track
of all symmetries of the effective Lagrangian. However, we think it is also 
instructive to
check how the effective perturbative potential \eq{oneloop} emerges
from  the Lagrangian \eq{LAGR}. The energy density $\theta^{00}$
corresponding to  \eq{LAGR} is given by  
\begin{eqnarray}
\theta^{00}(x)&=&
\frac{\vac}{2m^2}\,[(\partial_0\chi)^2+(\partial_i\chi)^2]\,e^{\chi/2}
-\mathrm{g}^{00}\,\vac\,e^\chi\,(1\,-\,\chi)\,
\nonumber\\
&& 
+\,\left(-F^{a0\lambda}F^{a0}_{\quad\lambda}\,+
\,\frac{1}{4}\,\mathrm{g}^{00}\,
(F_{\lambda\sigma}^a)^2\right)\,\left(
1\,-\,e^\chi\,(1\,-\,\chi)\right),
\end{eqnarray}
where $i=1,2,3$. Therefore the effective potential $W$ in the constant
chromomagnetic field $H$ is 
\beq\label{W}
W=
\int d^3x\,\left\{\frac{1}{2}\,H^2\,-\,e^\chi\,(1\,-\,\chi)\,
\left(\frac{1}{2}\,H^2\,+\,\vac\right)\right\}.
\eeq
In strong chromomagnetic field $H^2\gg \vac$ the energy density $W$
reduces to the effective potential \eq{oneloop}. In this case $\chi$
is not an independent degree of freedom, but rather a function of $H$.
 We calculate the
corresponding momentum scale in the next section.
The minimum of the functional $W(H,\chi)$ is found from the following
equations
\beq\label{min}
1\,-\,e^\chi\,(1\,-\,\chi) \,=\, 0, \quad
\chi\,e^\chi\,\left(\frac{1}{2}\,H^2+\,\vac\right)\,=\,0.
\eeq
Evidently, the minimum is at $\chi=0$ and the value of the
$W$ at the 
minimum is $-\vac$  independently of the value of
the chromomagnetic field $H$. We conclude that the physical vacuum of
the gluodynamics is described by one scalar field even in the presence
of the applied chromomagnetic field.  This justifies our assumption 
\eq{split}.

It is seen from \eq{W} that an increase of the color field $H$ leads to 
the increase of the energy density of the system. Since $H\sim g/r^2$ 
(where $r$ 
is the size of the system) the energy density decreases with $r$. At the 
same time the volume which the system occupies increases as $r^3$. 
Therefore, we 
expect that there exists a static configuration with a finite size $r_0$ 
such 
that the total energy of the system in minimal. This is analogous 
to the mechanism of bag formation in the Friedberg-Lee model \cite{FL}. 
However, the minimum of the effective potential in our model is 
independent of $H$ and located at $\chi=0$, while in the Friedberg-Lee 
model it depends on the density of the color sources.

Note that we can read the one-loop behavior of the strong coupling
right off the expression \eq{oneloop} for 
the effective potential. Indeed, the susceptibility of the vacuum in
the strong external chromomagnetic field is (do not confuse $\mu(H)$ with 
the renormalization scale $\mu$ in \eq{oneloop})
\beq\label{suc}
\mu(H)\,=\, 1\,-\,\frac{\beta(g)}{g}\,\left(
\ln\frac{g\,H}{\mu^2}\,-\,\frac{1}{2}\right)
\eeq
Recall that the beta function can be interpreted as a response of the system to the change of the external field. Namely, 
\eq{suc} implies  
\beq\label{beta2}
\beta=-g\,\frac{\partial\mu(H)}{\partial\ln H}
\eeq 
From \eq{LAGR} it follows that $\mu(H)$ is independent of $H$ at long 
distances, therefore $\beta=0$. The strong coupling does not run if
the effective theory is considered at the tree level. We will see in the
next section that quantum corrections do not alter that conclusion. 

It remains to check that the vacuum at $\chi=0$ is stable. Let us recall
that \eq{oneloop} is the real part of the perturbative effective
potential. However the perturbative potential has also the imaginary
part, as discussed above, which is due to the instability of the Landau level 
with $n=0$ and $s_z=1$, i.e.\ spin direction is parallel to the field.

Let us now examine the properties of the Landau levels in our effective theory near the $\chi=0$. The equation of
motion of the dilaton field is
\beq\label{EQMOT}
\frac{\vac}{m^2}\,\partial_\mu\left(
  e^{\chi/2}\,\partial_\mu\chi\right)
\,-\,
\frac{\vac}{4m^2}\,e^{\chi/2}\left(\partial_\mu\chi\right)^2\, +\,
\chi\, e^\chi\,\vac\,
+\,
\chi\, e^\chi\,\frac{1}{4}\,F_{\mu\nu}^aF^{a\mu\nu}\,
=\,0.
\eeq
Expanding near the minimum we arrive at 
\beq\label{effmass}
\partial_\mu^2\,\chi\,+\,m^2\,\left(1+\frac{H^2}{2\,\vac}\right)\,\chi
\,=\,0.
\eeq
The corresponding Landau levels are 
\beq\label{levels}
p^\mu\,p_\mu\,=\,m^2\,\left(1+\frac{H^2}{2\,\vac}\right).
\eeq
It is seen that $p_\mu p^\mu\ge 0$ for any $H$ so that 
the instability does not develop in the effective
theory we are discussing in this paper.
 
Next, consider the trace of energy-momentum 
tensor which can be  calculated directly from  \eq{LAGR} using  
\beq\label{TRACE}
\theta_\mu^\mu\,=\,\delta^{\mu\nu}\,\left(
 2\,\frac{\partial\lag}{\partial\delta^{\mu\nu}}\,-\,
\delta^{\mu\nu}\,\lag\right)
\,+\,
\frac{8\vac}{m^2}\,\partial_\mu^2\, e^{\chi/2},
\eeq
where the last term in the right hand side is the total
derivative. Using equation of motion of the dilaton field \eq{EQMOT}
one arrives at 
\beq\label{SP}
\theta_\mu^\mu\,=\,-\,4\,\vac\, e^\chi\,-\,\chi\, e^\chi\, F_{\mu\nu}^a\,
F^{a\mu\nu}.
\eeq
By virtue of \eq{EQMOT} one can clearly see that in the limit
$\vac\rightarrow 0$  the trace \eq{SP} vanishes and the classical
symmetries of the Yang-Mills Lagrangian are restored.
One might be worried that in the expansion of \eq{SP} in powers of $\chi$
the term $\chi F_{\mu\nu}^aF^{a\mu\nu}$ appears while 
it is absent in the Lagrangian \eq{LAGR}. However it is easy to see that this term is
canceled  out by the pure dilaton contribution. Indeed,
expanding the equation of motion \eq{EQMOT} up to the quadratic terms
in $\chi$ one finds
\beq\label{sokr}
\frac{1}{4}\,F_{\mu\nu}^a\, F^{a\mu\nu}\,\chi\,=\,-\,\vac\,\chi\,+
\, \mathrm{full\;\, derivative}\,+\,\order{\chi^2}\,.
\eeq

It is important to check that $\theta_\mu^\mu$ satisfies the low energy
theorems \eq{LET}. We have noted just after \eq{LAGR} that
$F_{\mu\nu}^a\sim q^2$. 
Hence it does not contribute to the $\theta_\mu^\mu$ at $q\rightarrow 0$. 
The only remaining term in \eq{SP} is the first one; it
saturates the low energy theorems as was explicitly proved in \cite{ms}.  
Thus, we have at our disposal the effective Lagrangian which (i) satisfies
the constraints imposed by  the Renormalization Group, \eq{SP};
(ii) the classical minimum $\chi=0$ of its  effective potential \eq{W}
is the true stable vacuum of QCD; (iii) its scale and conformal symmetry is
restored if $\vac =0$. 

\section{Quantum fluctuations around the physical vacuum}

Effective theory \eq{LAGR} is non-renormalizable. Let $M_0$ be its UV 
cutoff (in the effective potential method, this is the scale which corresponds to the lowest Landau level). 
Quantum fluctuations can develop only if there is enough kinematical 
space which is the case if  $m\ll M$. Let us define the 
perturbative expansion parameter $\lambda$ as
\beq\label{lambda}
\lambda\,=\, \frac{m}{M_0}\,.
\eeq
We will see later in this section that perturbative series in powers of
$\lambda$ is equivalent to the expansion of the Lagrangian \eq{LAGR} in
powers of $\chi$, and $\lambda$ indeed is the small expansion parameter in our 
effective theory.
For the rest of this section we assume that $\lambda$ is small and prove
this assumption in section~\ref{sec4}.

\subsection{Normalization of the energy-momentum tensor}

Let us first find the scale $M_0$ at which our effective
description breaks down; we will work in the leading order in $\lambda$.
The vacuum expectation value of the trace of
energy-momentum tensor \eq{SP} is the physical observable and does not
depend on a particular choice of degrees of freedom in the
Lagrangian; its value is given by \eq{treb}.
By virtue of \eq{SP} it is equivalent to the requirement that
\beq\label{req}
4\,\vac\,\langle \, 1\,-\,e^\chi\rangle\,=\, 
\langle \chi\, e^\chi\,F_{\mu\nu}^aF^{a\mu\nu}
\rangle .
\eeq
In the vacuum $\chi=0$ \eq{treb} is obviously
satisfied. Quantum fluctuations in general violate
this requirement. However, since the 
effective Lagrangian \eq{LAGR} is formally divergent at
short distances we have to impose an ultra-violet cutoff $M_0$. 
We will choose such a cutoff that \eq{treb} is satisfied.

Expanding \eq{SP} to the order $\order{\chi^0}$ we  
obtain a trivial result  
\beq\label{uxty}
\theta_\mu^\mu\,=\,-4\,\vac\,+\,\order{\chi}.
\eeq
At the order $\order{\chi}$ Eq.~\eq{req} is  satisfied due to 
\eq{sokr}. 
\begin{figure}
\begin{center}
\epsfig{file=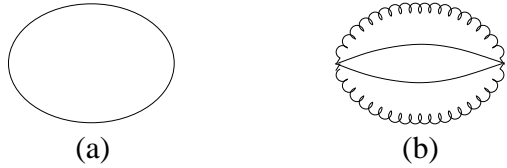,width=7cm}
\caption{\sl (a) Pure dilaton contribution to the trace of 
energy-momentum tensor; (b) mixed dilaton-gluon  contribution to the
two-point correlator of the trace of energy-momentum tensor.
}\label{f:bubles}
\end{center}
\end{figure}
At the next order $\order{\chi^2}$ \eq{req} can be satisfied 
only for a particular choice of the cutoff $M_0$.
 Note that by the low energy theorems \eq{LET} (with $n=1$)
long distance
contributions to expectation value of the operator
$\theta_\mu^\mu(x)$ can be expressed through the two point correlator
$\Xi(q^2)$ defined as
\beq\label{xi}
\Xi(q^2)\,=\,i\,\int\,d^4x\, e^{i\,q\,x}\,\langle\,0\,|\,T\,
\theta_\mu^\mu(x)\theta_\mu^\mu(0)\,|\,0\,\rangle\,=\,
\int\,d\sigma^2\,\frac{\rho(\sigma^2)}{\sigma^2\,-\,q^2\,-i0}\,,
\eeq
where we have introduced the spectral density $\rho(q^2)$.
We find it more convenient to work with this correlator. The first
reason is that the spectral density can be expressed in terms of
physical states. The other one is that we know $\rho(q^2)$ for the
perturbative theory. 

To rewrite condition \eq{req} in terms of two-point correlator
we apply to it the low energy theorem \eq{LET}
\beq\label{kju}
\Xi(0)\,=\,-4\,\langle\,\theta_\mu^\mu\,\rangle\,.
\eeq
Thus, our requirement that \eq{req} holds at the leading non-trivial order
in $\lambda$  can be written as (see \eq{SP})
\beq\label{blin}
\Xi_\mathrm{dil}(0)\,+\,\Xi_\mathrm{mix}(0)\,=\,0\,,
\eeq
where we separated the pure dilaton and mixed dilaton-gluon contributions.

The pure dilaton contribution can be read from \eq{SP}:
\beq\label{pdlo}
\langle\,\theta_\mu^\mu\,\rangle_\mathrm{dil}\,=\,
-4\,\vac\,\frac{1}{2}\,\langle\,\chi^2\,\rangle
\eeq
which implies that (see \fig{f:bubles}(a))
\beq\label{oli}
\Xi_\mathrm{dil}(0)\,=\, 8\,\vac\, \langle\,\chi^2\,\rangle\,=\, 
8\,\vac\,\frac{m^2}{\vac}\,\frac{1}{2}\,
\int\,\frac{d^4k}{(2\pi)^4}\,\frac{i}{k^2}\,=\,
\frac{m^2\,M_0^2}{2\,\pi^2}\,.
\eeq

Let us turn to the mixed gluon-dilaton contributions. 
The corresponding diagram is shown in \fig{f:bubles}(b). Its imaginary
part is  
\begin{eqnarray}
\rho_\mathrm{mix}(\sigma^2)&=&
\left(\frac{m^2}{\vac}\right)^2\, \frac{N_c^2\,-\,1}{4}\,
\int\,\frac{d^4q_1}{(2\pi)^4}\,\int\,\frac{d^4q_2}{(2\pi)^4}\,
(q_1^\mu q_2^\nu\,-\, (q_1\cdot q_2)\,g^{\mu\nu})^2
(2\pi)^2\,\delta(q_1^2)\,\delta(q_2^2)\,
\nonumber\\
&&\times\,
\int\,\frac{d^4k_1}{(2\pi)^4}\,\int\,\frac{d^4k_2}{(2\pi)^4}\,
(2\pi)^2\delta(k_1^2)\,\delta(k_2^2)\,(2\pi)^4\,\delta(k_1+k_2+
q_1+q_2)\nonumber\\
&=& 
\frac{\sigma^{8}}{140\cdot 48\,(2\pi)^5}(N_c^2-1)\,\left(\frac{m^2}{\vac}
\right)^2,
\label{manda}
\end{eqnarray}
where we neglected the mass of the dilaton $m$ with respect to the
cutoff $M_0$.
$\Xi(q^2)$ can be calculated using dispersion relation with
subtractions
\begin{eqnarray}
\Xi(q^2)_\mathrm{mix} &=&
\int_0^\infty\frac{\rho_\mathrm{mix}(\sigma^2)\,d\sigma^2}
{\sigma^2 \,-\,q^2\,-\,i0}\,-\, 
q^2\,\int_0^\infty\frac{\rho_\mathrm{mix}(\sigma^2)\,d\sigma^2}
{\sigma^4}\nonumber\\
&&-\, 
q^4\,\int_0^\infty\frac{\rho_\mathrm{mix}(\sigma^2)\,d\sigma^2}
{\sigma^6}\,-\,\ldots\,-\,
q^{12}\,\int_0^\infty\frac{\rho_\mathrm{mix}(\sigma^2)\,d\sigma^2}
{\sigma^{14}}\nonumber\\
&=& 
q^{10}
\,\int_0^\infty\frac{\rho_\mathrm{mix}(\sigma^2)\,d\sigma^2}
{\sigma^{10}\,(\sigma^2 \,-\,q^2\,-\,i0)}\,-\, 
\int_0^\infty\frac{\rho_\mathrm{mix}(\sigma^2)\,d\sigma^2}
{\sigma^2}\,.
\label{regco}
\end{eqnarray}
The dispersion integral in the last line of \eq{regco} is proportional to 
$q^{10}\ln(-M_0^2+q^2)$. Consequently,
\beq\label{mixed}
\Xi_\mathrm{mix}(0)\,=-\, 
\int_0^\infty\frac{\rho_\mathrm{mix}(\sigma^2)\,d\sigma^2}
{\sigma^2}
\,=\,-\,\frac{1}{4}\,\rho_\mathrm{mix}(M_0^2).
\eeq
Formally, \eq{mixed} gives the value of the non-vanishing subtraction
constant in the dispersion relation.

Substituting  \eq{oli} and \eq{mixed} into vacuum stability
condition \eq{blin} results in the equation determining the
ultra-violet cutoff $M_0$ of the effective theory \cite{klt}
\beq\label{stab}
M_0^2\,=\, 16\,\pi\,105^{1/3}\,(N_c^2\,-\,1)^{-1/3}\,
\left(\frac{\vac}{m}\right)^{2/3}.
\eeq

\subsection{Gluon polarization tensor}

\begin{figure}
\begin{center}
\epsfig{file=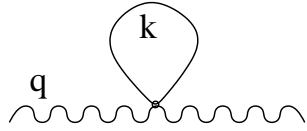,width=4cm}
\caption{\sl The leading order  quantum correction to the
  gluon propagator at long distances.
}\label{f:tadpole}
\end{center}
\end{figure}

We have argued that the vacuum expectation of
the gluon condensate \eq{treb} is unchanged provided we had chosen 
the value of the cut-off according to \eq{stab}. In that case  
the quantum corrections does not
change the vacuum energy density which is completely saturated by
classical solution. Now we would like to calculate quantum corrections
to the strong coupling. To the leading order in $\lambda$
we have the tadpole diagram in \fig{f:tadpole}
Introduce the scalar function $\Pi(q^2)$ as follows
\beq\label{pdef}
\Pi_{\mu\nu}(q)\,=\,(q^\mu\, q^\nu\,-\, q^2\, g^{\mu\nu})\,\Pi(q).
\eeq
The tadpole diagram is given by 
\beq
i\,\Pi^\mathrm{tadpole}_{\mu\nu}(q)\,=\,\frac{1}{2}\,
\frac{m^2}{\vac}\,\int_m^{M_0}\,\frac{d^4
  k}{(2\pi)^4}\,\frac{i}{k^2-m^2+i0}\,i\,(-1)(q^\mu\, q^\nu\,-\,
q^2\, g^{\mu\nu})\,.
\eeq
It can be calculated by performing
Wick rotation and consequent integration over a four dimensional
sphere of radius $M_0$. We neglect
then the dilaton mass which gives contribution of higher order in
$\lambda$.
 The result of calculation of the diagram in \fig{f:tadpole}  is
\beq\label{tadpole}
\Pi(q^2)\,=\, \frac{M_0^4}{64\pi^2\,\vac},
\quad q^2\,\le\, M_0^2.
\eeq
The quantum correction is constant. This means that the strong
coupling freezes at long distances.

The tadpole diagram \fig{f:tadpole} is leading order in $\lambda$
correction. However the  higher order corrections deserve a special
remark since they could be in principle logarithmically divergent at 
$q^2=4m^2$ in which case those diagrams would dominate the polarization
tensor at long distances. In the Appendix~B we  argue that all such 
logarithms appear in the product $(q^2-4m^2)^n \ln(q^2/4 m^2)$, where  
$n>0$  and thus vanish at the end-point of the dilaton spectrum.
Also we check that the sub-leading diagrams  are numerically
small. Therefore, the conclusion of the previous section that the
strong coupling freezes at long distances holds if such quantum
corrections are included.

Eq.~\eq{tadpole} is the leading order contribution of vacuum
fluctuations to the gluon polarization tensor. We can systematically
develop the perturbation theory in $\lambda$. 
The qualitative picture of the renormalization group
flow can be obtained by simple dimensional analysis
(see e.g.\ \cite{ps}). Since the typical scale for mass is the cutoff
$M_0$, the coefficients in front of the four terms 
in the rhs of \eq{LAGR} have the following behavior at different
momentum scales $p$: $(M_0/p)^2$, $ (M_0/p)^4$, $ (M_0/p)^0$ and 
$(M_0/p)^2$ 
respectively.  Thus, the only relevant term at low momenta is the
second one, which is purely dilatonic term. This is a
manifestation of the fact that the dynamics of the vacuum fields
decouples from the colored sources.

\section{Matching onto the perturbation theory}\label{sec4}

One can express the strong coupling at the
cutoff $M_0$ as  function of the parameters of the low energy Lagrangian.
This can be achieved by
matching the spectral density of the effective theory \eq{manda} with
the spectral density of perturbation theory  at $M_0^2$. In
perturbative gluodynamics the anomalous trace of energy-momentum
tensor is given by \eq{anom}.
Then the calculation of the spectral density of the 
correlator \eq{xi} is straightforward \cite{fh}
\begin{eqnarray}
\rho_\mathrm{pert}(q^2)&=& \left(\frac{b\,\alpha_s}{8\,\pi}\right)^2\,
4^2\,\frac{N_c^2\,-\,1}{2}\,
\int\,\frac{d^4q_1}{(2\pi)^4}\,\int\,\frac{d^4q_2}{(2\pi)^4}\,
(q_1^\mu q_2^\nu\,-\, (q_1\cdot q_2)\,g^{\mu\nu})^2\nonumber\\
&& \times\,(2\pi)^2\,\delta(q_1^2)\,\delta(q_2^2)\, (2\pi)^4\,\delta(q+
q_1+q_2)\nonumber\\
&=& 
 \left(\frac{b\,\alpha_s}{8\,\pi}\right)^2\,\frac{(N_c^2-1)}{2\,\pi}\,q^4.
\label{pert}
\end{eqnarray}
 Since the spectral density is just the
imaginary part of the correlator, it is clear that only mixed diagrams
of \eq{mixed} contribute to the matching in the leading in $\lambda$ order
(indeed,  $\rho_\mathrm{dil}\sim m^4$)
\beq\label{vot}
\rho_\mathrm{mix}(M_0^2)\,=\,\rho_\mathrm{pert}(M_0^2)\,.
\eeq
Using Eq.~\eq{stab} we obtain
\beq\label{match}
\alpha_s(M_0^2)\,=\,\frac{16\,\sqrt{\pi}\,\lambda}{b\,\sqrt{N_c^2\,-\,1}}
,\quad -q^2\,=\, Q^2\le M_0^2,
\eeq
This equation shows that the small parameter of perturbation theory $\alpha_s$  
is matched onto the small parameter of our effective theory, $\lambda$.

At $Q^2> M_0^2$ the strong coupling runs as
\beq\label{gene}
\alpha_s(Q^2)\,=\, \frac{\alpha_s(M_0^2)}{\left(1\,
+\, \frac{b\,\alpha_s}{4\,\pi}\,\ln\frac{Q^2}{M_0^2}
\right)}\,=\,
\frac{4\,\pi}
{b\,\ln\frac{Q^2}{\Lambda_\mathrm{QCD}^2}},\quad Q^2> M_0^2,
\eeq
where we introduced the familiar phenomenological constant 
constant $\Lambda_\mathrm{QCD}$ as 
\beq\label{lam}
\Lambda_\mathrm{QCD}^2\,=\, M_0^2\,
e^{-\frac{4\,\pi}{b\,\alpha_s(M_0^2)}}.
\eeq

\subsection{Numerical estimations}

QCD sum rules analysis performed in \cite{svz,svz1} make it possible
to estimate the non-perturbative scale inherent to the vacuum of gluodynamics, which appears quite hard:  
$M_0^2=20$~GeV$^2$. Lattice calculations show \cite{lattice} that the 
lightest  resonance   in pure gluodynamics is the scalar glueball with mass
$m \simeq 1.6$~GeV. It is natural to identify this glueball with dilaton; it is interesting 
that this state appears to have a size much smaller than the sizes of 
glueballs with other quantum numbers \cite{lattice}. In the approach followed in this paper 
this is a consequence of a large value of the cutoff scale $M_0$. 
From the vacuum stability condition \eq{stab} we 
find $\vac\simeq(0.58~\mathrm{GeV})^4$. By definition
$\lambda=m/M_0\simeq 0.36$. 
Eq.~\eq{match} then implies that $\alpha_s(M_0^2)\simeq 0.33$. The
value of the $\Lambda_\mathrm{QCD}$ follows from \eq{lam}:
$\Lambda_\mathrm{QCD}\simeq 0.79$~GeV.

In the world with light quarks the scalar glueball mixes with scalar $\bar{q}q$ meson \cite{sa}. The lightest scalar resonance is the
$\sigma$-resonance which is a strong mixture of the glueball and the $q\bar
q$ meson \cite{efk}. 
In this case the dilaton mass can be estimated as
the mass of the $\sigma$ \cite{klt}: $m\simeq 0.6$~GeV.
QCD sum rules give an estimate of the QCD vacuum energy density:
$\vac=(0.24~\mathrm{GeV})^4$. From \eq{stab} we have $M_0\simeq
1.9$~GeV. Other estimates can be done exactly as in the previous
paragraph yielding $\lambda=0.31$, $\alpha_s(M_0^2)\simeq 0.35$ and 
$\Lambda_\mathrm{QCD}=0.26$~GeV. To verify how good is this
value from the phenomenological point of view we use \eq{gene} and
find that at the Z-boson mass scale $\alpha_s(m_Z)\simeq 0.12$.
This is in a reasonable agreement with the data -- see 
discussion in Ref.~\cite{revs}.

\subsection{Dependence on $N_c$}

\begin{figure}
\begin{center}
\epsfig{file=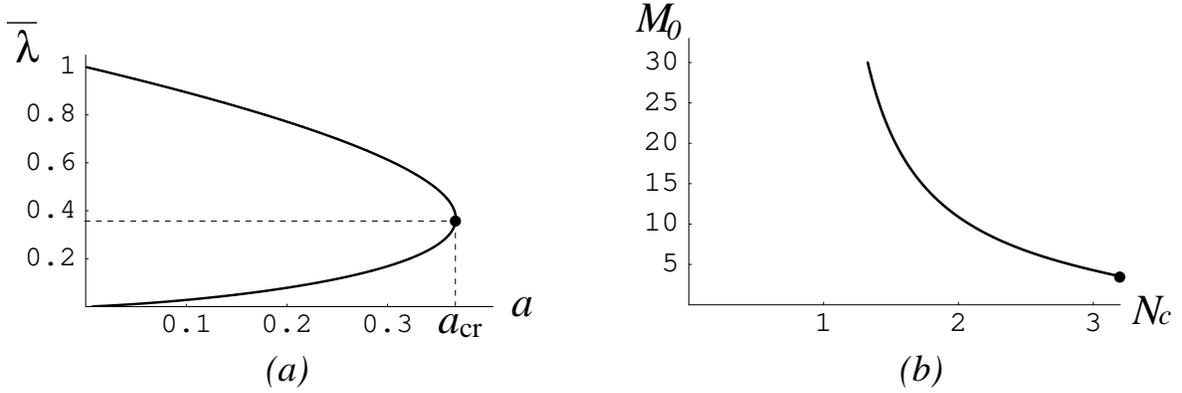,width=15.5cm}
\end{center}
\caption{\sl Numerical solution to \eq{crit}. (a) Rescaled coupling
  $\bar\lambda$ vs $a$; the critical point is $(e^{-1},e^{-1})$.  
 (b) Dependence of $M_0$ on the number of colors in
 gluodynamics ($\Lambda_\mathrm{QCD}=0.8$~GeV); 
only the  physical branch of the solution is shown. With a good
accuracy  $M_0\propto 1/N_c^2$.}\label{f:crit}
\end{figure}

Let us now discuss the dependence of our effective theory on the number of
colors $N_c$.
Eq.~\eq{lam} can be considered as an equation for the cutoff of the
effective theory $M_0$ as a function of the number of colors
$N_c$. Let us find  $M_0(N_c)$. It is convenient to
introduce the dimensionless parameter $a$ and have $\lambda$ rescaled 
as follows
\beq\label{resc}
\bar\lambda\,=\,
\frac{\lambda\,\Lambda_\mathrm{QCD}}{m},\quad
a\,=\,\sqrt{N_c^2\,-\,1}\,\frac{\Lambda_\mathrm{QCD}\sqrt{\pi} }{8\,m}.
\eeq
Then using \eq{lambda} and \eq{resc} Eq.~\eq{lam} takes form
\beq\label{crit}
\bar\lambda\,=\, \exp\left\{-\frac{a}{\bar\lambda}\right\}\, . 
\eeq
Its solution is shown in \fig{f:crit}(a). We observe that the solution has
two branches: one starting at the origin $(0,0)$ and the second one starting at
the point $(0,1)$. Both branches terminate at the critical point
$(a_\mathrm{cr},\bar\lambda_\mathrm{cr})=(e^{-1},e^{-1})$. To
pick up the physical branch we note that by \eq{resc} $a=0$ when
$N_c=1$. Thus, by \eq{match} $\bar\lambda =0$ at this point. Therefore
the physical branch is the lower one in \fig{f:crit}(a). In
\fig{f:crit}(b) we represent it as a plot of the cutoff $M_0$ versus
the number of colors $N_c$. The value of $\Lambda_\mathrm{QCD}=0.8$ chosen
for this figure is such that $\alpha_s(M_Z^2)=0.12$ at $b=11$. 
The critical value of the parameter $a$ corresponds to the critical
value of $N_c$.  Using \eq{resc} we find
\beq\label{oxr}
N_c^2\,\le\, N_c^{\mathrm{cr}\,2}\,=\,
\left(\frac{8\,m}{\sqrt{\pi}\,\Lambda_\mathrm{QCD}}
\,e^{-1}\right)^2\,+\,1,
\eeq
(of course, the integer part of the right-hand-side must be taken)
which yields $N_c^\mathrm{cr}=3$. When $N_c>N_c^\mathrm{cr}$ our effective
theory ceases to be valid. Indeed, $M_0$ rapidly decreases
 with $N_c$ (approximately as $1/N_c^2$) approaching the dilaton mass $m$.   
The values  of the strong coupling
$\alpha_s(M_0^\mathrm{cr})$ and the cutoff $M_0^\mathrm{cr}$ at the
critical point are 
\beq\label{fin}
\alpha_s(M_0^\mathrm{cr})\,=\,\frac{2\,\pi}{b^\mathrm{cr}}\quad,
\eeq
and
\beq\label{fin2}
M_0^\mathrm{cr}\,=\,\frac{8\, m}{\sqrt{\pi(N_c^\mathrm{cr\, 2}-1)}}.
\eeq

We see that the effective theory breaks down at large $N_c$, with 
the critical value $N_c^\mathrm{cr}$ (see \eq{oxr}). 
This can be readily interpreted if we
recall that the perturbative vacuum energy density
grows as $\vac^\mathrm{pert}\sim N_c^2$ at large $N_c$, whereas $\vac$
of the effective theory does not  (see
\fig{f:crit}(b) and \eq{stab}). Thus, the effective
theory breaks down at such large values of $N_c$
 that the perturbative vacuum energy density cannot be matched 
onto the effective one.  In the region $N_c\le N_c^\mathrm{cr}$ where we 
can use 
the effective theory, the $N_c$
dependence of the value of the freezing strong coupling is given by 
$\alpha_s(M_0^2)b\sim\lambda/N_c\sim N_c$ as can be seen from
\eq{match} and \fig{f:crit}(b).  

Since $M_0$ decreases as $N_c$ increases the matching region is driven
into the infrared where the perturbative expansion can no longer be trusted.  
Indeed, the dilaton spectral density vanishes at $q^2<4m^2$. 
On the contrary, the perturbative gluon spectral density is finite at
arbitrary small but finite $q^2$ (see \eq{pert}). Although the dilaton
effective theory takes into account the non-perturbative effects
associated with the scale anomaly, it is not clear how 
those effects are  related to the color potential at long distances. The
interplay between the dilaton low energy effective theory 
 and the gluodynamics at large $N_c$ certainly deserves
special study. 


\section{Conclusions}

In this paper we constructed an effective low energy Lagrangian \eq{LAGR} 
of gluodynamics which 
(i) satisfies the constraints imposed by the Renormalization Group; (ii) 
its vacuum is stable; (iii) it is scale and 
conformally invariant in the limit of vanishing 
 vacuum energy density $\vac$; (iv) it matches onto 
the perturbative theory at short distances. 
Using this Lagrangian we 
 developed the perturbation theory of quantum fluctuations around the 
physical vacuum. Since the effective theory \eq{LAGR} is divergent when
considered on a quantum level we must introduce an ultraviolet
cutoff $M_0$. To calculate it we noted that classical configuration of
the dilaton field saturates the vacuum energy $\vac$. Therefore the value of $M_0$ is dictated by 
the requirement of vacuum stability -- quantum fluctuations
must not contribute to the vacuum energy density. This happens to be
true only for a certain choice of $M_0$ given by \eq{stab}. In the
kinematic region $q^2\le M_0^2$ we  developed a perturbation theory
in a small parameter $\lambda=m/M_0$ and used it to calculate the
leading \eq{tadpole} and next-to-leading \eq{sublead} order 
radiative corrections to the gluon propagator.
We observed that the leading radiative correction to the gluon propagator
is constant.  We conclude that  the strong coupling $\as$  freezes 
at distances larger than the inverse cutoff $1/M_0$; this behavior is consistent with 
the analysis of  Refs.~\cite{revs},\cite{brodsky} .

By matching the spectral densities of the perturbation theory 
and of the effective one we determined the value 
of the strong coupling at the scale $M_0$ in terms of the 
vacuum energy density $\vac$ and the glueball mass $m$, \eq{match}.
Using QCD sum rules to estimate $M_0^2\simeq 20$~GeV$^2$ we calculate the
$\Lambda_\mathrm{QCD}$ and then $\alpha_s(m_Z)$; we found a reasonable
agreement with experimental data. We consider this as an
additional evidence that the typical scale of vacuum fluctuations of
QCD is hard \cite{svz1,fh,shuryak}. 

We discussed the $N_c$ dependence of the theory. As $N_c$ increases
$M_0$ decreases as $\sim 1/N_c^2$, so that at some $N_c^\mathrm{cr}$
we have $M_0\le m$ and the quantum fluctuations of the dilaton field
are no longer possible. The matching on the perturbation theory
\eq{vot} and \eq{match} breaks down. Numerically, $N_c^\mathrm{cr}$ is
found to be just above 3, so the effective theory \eq{LAGR} is applicable to
the study the infrared behavior of $SU(3)$ gluodynamics.

One of our main results - the freezing of the strong coupling at long distances -- 
has an elegant geometric interpretation. Recall that we derived the 
effective Lagrangian \eq{LAGR} by formally coupling Yang-Mills theory to the
conformally flat gravity described by the field $\chi$ \cite{ms}. 
This way the scale symmetry of Yang-Mills theory is restored at the cost of
introducing a new field. At very short distances $\as\ll 1$ and the
scale anomaly vanishes in usual perturbative  gluodynamics.
Effectively this means considering Yang-Mills  theory 
in the flat space. At very long distances the theory resides in its physical 
vacuum $\chi=0$, see \eq{riemann}, which means that the spacetime is flat
again. In between those extreme cases we can think of  Yang-Mills 
field as a classical field propagating on a curved background. 
Indeed it has been found  \re{grav} that the coupling of the
Yang-Mills theory on a curved background freezes at long distances.
 
The physical picture which has emerged from our study thus corresponds
to color fields dynamically confined within a cavity by the interaction with 
self-coupled scalar glueball fields. This interaction regularizes the theory 
in the infrared region, and leads to the freezing of strong coupling at large 
distances. 
 
It will be very interesting to study the properties of bound states in this  
 ``conformal bag model".  While we checked that the model does have 
the corresponding solutions, so far we have not succeeded in finding 
them analytically.  

A crucial test of the ideas presented in this paper can be performed on the 
lattice. Since $r_0 \sim 1/M_0$ corresponds to the size of the scalar glueball, and $M_0$ 
decreases as a function of $N_c$, we predict that the scalar glueball in $SU(4)$ gauge 
theory will have a larger size than in $SU(3)$. Unlike in $SU(3)$ theory, where the scalar glueball was found 
to have the smallest size (see e.g. \cite{lattice}), in $SU(4)$ we expect all glueballs to have similar sizes. 
 In contrast, in $SU(2)$ theory the size of the 
scalar glueball should become even smaller than in $SU(3)$. These predictions can be tested directly  
by measuring the glueball formfactors (three-point correlation functions), or indirectly by measuring the two-point 
correlation functions of the scalar gluon operators and by checking at what distances they approach 
the perturbative behavior. If the lattice results in gluodynamics confirm the validity of the effective theory advocated in this paper, 
  it will be worthwhile to include the light quarks by putting the classical QCD Lagrangian on the curved background. 
  This could then substitute a consistent theoretical approach to the study of infrared behavior in QCD.  

\acknowledgments
We acknowledge  interesting discussions on the subject with A.~Gotsman,
Yu.~Kovchegov, U.~Maor, M. Praszalowicz  and D.T.~Son. We are indebted to 
T.D. Lee 
for valuable comments and suggestions, and to H.~Meyer and P.~Petreczky for a  
discussion of the lattice data on glueballs. 
The work of D.K. and K.T. was supported by the U.S. Department of Energy 
under Contract No. DE-AC02-98CH10886.
This research was supported in part by the GIF grant \# I-620-22.14/1999 
and by Israeli Science Foundation, founded
by the Israeli Academy of Science and Humanity. At the early stage the 
work of K.T. was sponsored  by the U.S. Department of Energy under 
Grant No. DE-FG03-00ER41132. 


\begin{appendix}

\begin{figure}
\begin{center}
\begin{tabular}{ll}

\epsfig{file=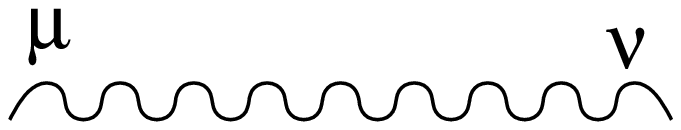,width=4cm}
& 
$D_g(q^2)\,=\,\frac{-i\,g^{\mu\nu}\,\delta^{ab}}{q^2\,-\,i0}$
\\[0.8cm]

\epsfig{file=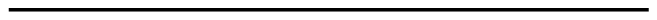,width=4cm}&
$D_d(k^2)\,=\, \frac{m^2}{\vac}\,\frac{i}{k^2\, -\, m^2\, -\, i0}$\\[0.5cm]

\begin{minipage}{4.1cm}
\epsfig{file=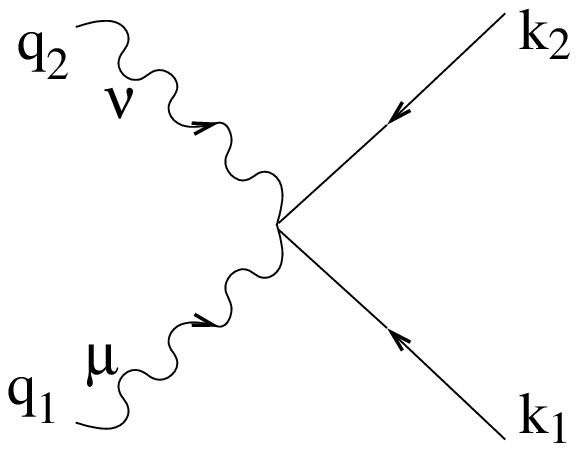,width=4cm}
\end{minipage}
&
\begin{minipage}{5cm}
$i\,
\left((q_1\cdot q_2)g^{\mu\nu}-q_1^\nu q_2^\mu\right)\,\delta^{ab}$
\end{minipage}

\end{tabular}
\end{center}
\caption{\sl Feynman rules for the dilaton effective theory.}\label{f:fey}
\end{figure}

\section{Feynman rules for the dilaton Lagrangian}
In Appendix~A we list the Feynman rules for the Lagrangian
\eq{LAGR} up to the quadratic terms in $\chi$, 
see \fig{f:fey}.
Here $a$ and $b$ are the color indexes.
We observe that dilaton graphs do not violate the color symmetry.
This is seen of course directly from the Lagrangian \eq{LAGR}.

\end{appendix}

\begin{appendix}
\setcounter{section}{1}
\section{Higher order corrections to the gluon polarization tensor}

\begin{figure}
\begin{center}
\epsfig{file=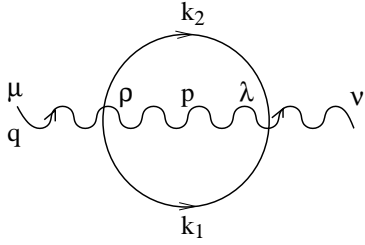,width=5.2cm}
\end{center}
\caption{\sl Next-to-leading order diagram contributing to the gluon
  polarization tensor.}\label{f:polar}
\end{figure}

In Appendix~B we argue that the higher order corrections to the
gluon polarization tensor have no singularities at the end-point of
the dilaton spectrum $q^2=4m^2$. Let us consider the diagram 
\fig{f:polar} for example.

We have 
\begin{eqnarray}
i\,\Pi_{\mu\nu}^b(q)&=&i\,\Pi_{\mu\nu}(q) \nonumber\\
&=&
\left(\frac{m^2}{\vac}\right)^2\,\frac{1}{2!}\int\,\frac{d^4p}{(2\pi)^4}
\,\frac{d^4k_1}{(2\pi)^4}\,\frac{d^4k_2}{(2\pi)^4}\,i\,
(p^\mu q^\rho -(qp)g^{\mu\rho})\,\frac{-i\,g^{\rho\lambda}}{p^2}\nonumber\\
&&
\times\,i\,
(p^\nu q^\lambda -(qp)g^{\nu\lambda})\,
\frac{i}{k_1^2-m^2}\,\frac{i}{k_2^2-m^2}\,
(2\pi)^4\, \delta(k_1+k_2+p-q)\,.\label{GD}
\end{eqnarray}
Contracting Lorentz indexes and averaging over directions of $p$ it
can be shown that $\Pi_{\mu\nu}(q)$ has the same transverse structure as 
displayed in \eq{pdef}. Making contractions in the definition
\eq{pdef} we arrive at 
\begin{eqnarray}
\mathrm{Im}\,\Pi(q) &=&\frac{1}{3q^2}\,\mathrm{Im}\,\Pi^\mu_\mu(q)\,
=\,\frac{1}{3q^2}\,\left(\frac{m^2}{\vac}\right)^2\,
\frac{1}{16\pi}\,\int\, d^4k\, \int\,\frac{d^4p}{(2\pi)^3}\,
\nonumber\\
&&
\times\,\delta(k+p-q)\,\delta(p^2)\,2\,(pq)^2\,
\sqrt{1-\frac{4m^2}{M^2}}\nonumber\\
&=&
\frac{1}{3q^2}\,\left(\frac{m^2}{\vac}\right)^2\,
\frac{1}{64\pi^3}\int\, dM^2\, |\vec k|\, \omega_p^2\, M_q
\,\sqrt{1-\frac{4m^2}{M^2}}\,,\label{PR1}
\end{eqnarray}
where $M_q^2$ is the external gluon virtuality. 
Denote $t=M_q^2$. It is easily seen that 
\beq
\omega_p\,=\,|\vec p|\,=\,|\vec k|\,=\, \frac{t-M^2}{2\sqrt{t}}\,.
\eeq
Integral in \eq{PR1} over $M^2$ in the range $4m^2\le M^2\le t$
can be easily done giving somewhat lengthy result. Near the end-point
of the spectrum the result of integration is
\beq\label{MNIM}
\mathrm{Im}\,\Pi(t)\,\approx\, 
\left(\frac{m^2}{\vac}\right)^2\,
\frac{(t\,-\,4\, m^2)^{9/2}}{140\, (6 \pi)^3\, m\, t^2}\,.
\eeq
The polarization tensor can be calculated 
using dispersion relation
\beq\label{DISP}
 \Pi(q)\,=\,(q^2\,-\,4m^2)^5\, \frac{1}{\pi}\int_{4m^2}^{M_0^2}\,
\frac{\mathrm{Im}\,\Pi(q)}{(t-q^2-i0)\, (t\,-\, 4m^2)^5}\,dt\,.
\eeq
Dispersion relation can be applied only
to a function which vanishes sufficiently fast 
at infinite radius in the complex plain
of $t$. Therefore we apply it to a function
$\mathrm{Im}\Pi(t)/(t-4m^2)^5$ instead of  $\mathrm{Im}\Pi(t)$.
This procedure corresponds to the subtractions
\beq\label{subtr}
\Pi(q^2)\,\rightarrow\, \Pi(q^2)\,-\, 
\sum_{l=0}^4\,\frac{1}{l!}\Pi^{(l)}(4m^2)\,(q^2\,-\,4m^2)^l\,.
\eeq
It follows from \eq{MNIM}, \eq{DISP} and \eq{subtr} that  
\beq\label{sublead}
\Pi(q)\,\propto\, (t\,-\,4\, m^2)^{9/2}\,
\rightarrow\, 0, \quad \mathrm{as}\;\, t\,\rightarrow\, 4m^2\,, 
\quad l\neq 0\,.
\eeq
The term  with $l=0$ is just the largest subtraction constant (cp.\ 
 \eq{regco}).

Therefore we can safely expand \eq{PR1} in powers of
$\lambda$. Integrating over $M^2$ and using dispersion relation
\eq{DISP} with $m=0$ we obtain 
\beq
\Pi(Q^2)\,=\, \frac{Q^4\, m^4}{\vac^2}\,\ln\frac{M_0^2\,+\,Q^2}{Q^2}\,
\frac{1}{24\, (4\pi)^4}\,+\, \mathrm{const}. 
\eeq
At $Q^2=M_0^2$ this contribution reaches its maximal value $\sim \lambda^2$  
and thus parametrically and numerically suppressed with respect to the
leading result \eq{tadpole}.

We can easily extend our argument to higher order diagrams. Indeed, the 
introduction of  additional dilaton lines can only  bring in a
factors of $M^2/m^2$ as can be seen from the gluon-dilaton
vertex in the Appendix~A 
and Fig.~1(b) of \re{klt} for dilaton self interactions.

\end{appendix}

\end{document}